\documentclass[aps,prd,twocolumn,eqsecnum,showpacs,amsmath]{revtex4}
\usepackage[dvips]{color,graphicx}
\usepackage{amsfonts}
\usepackage{amssymb}
\usepackage{latexsym}

\newcommand{\dalm}{\kern1pt\vbox{\hrule height 0.9pt\hbox{\vrule width
0.9pt\hskip 2.5pt\vbox{\vskip 5.5pt}\hskip 3pt\vrule width 0.3pt}\hrule height
0.3pt}\kern1pt}
\newcommand{\ma}[1]{\mbox{$\mathcal{#1}$}}

\newcommand{\lw}[1]{\smash{\lower2.ex\hbox{#1}}}
\newtheorem{The}{Theorem}

\begin{document}


\title{Formation of a black hole from an AdS spacetime} 
\author{
 Naresh Dadhich\footnote{Electronic address:nkd@iucaa.ernet.in}}

\affiliation{
 Inter-University Centre for Astronomy \& Astrophysics, Post Bag 4, Pune~411~007, India\\
}
\date{\today}

\begin{abstract}
We shall first discuss motivation for higher dimension even for classical 
description of gravitational dynamics and then construct a black hole out of 
an anti-deSitter (AdS) spacetime by prescribing a coupling between Gauss-Bonnet parameter, constant curvature of extra dimensional space and $\Lambda$. This is a creation of pure curvature which establishes the fundamental reciprocity between matter and gravity/curvature.
\end{abstract}


\maketitle

\section{Introduction}

Einstein's theory of gravitation is uniquely distinguished from all other forces by its universal character - it links to all that physically exists. It is precisely for this reason that it could only be described by nothing else than spacetime curvature. That is why what is generally true for other fields may not be true for gravity. For example, one never asks the question that does a field remain confined to the three space dimensions or does it propagate in extra dimension? For gravity, this question is quite valid simply because it is equivalent to asking, does spacetime curvature remain confined to a given dimension? This means that even at the classical level, there is an issue of higher dimension for gravity and we shall first discuss three different classical motivations for higher dimension. 

Once in higher dimension, the proper action for gravitation should also include higher order curvature terms given by Lovelock polynomial \cite{lov}. This is the unique Lagrangian which leads to the quasi-linear (highest order of derivative being linear) second order differential equation. Also it is the only one for which  both Palatini and metric formulations give the same equation of motion \cite{jabb}. In this Lagrangian, $\Lambda$, Einstein-Hilbert and Gauss-Bonnet are respectively zeroth, linear and quadratic terms. For dimension less than five, the quadratic and higher order terms make no contribution in the equation and hence upto dimension four, $R$ and $\Lambda$ suffice. It turns out that contribution of Gauss-Bonnet term weakens the higher dimensional black hole singularity as the metric and gravitational field remain finite and regular everywhere \cite{boul,dad}, and Riemann curvature diverges weakly as $1/r^3$ rather than $1/r^4$ for the Schwarzschild particle in five dimension. The key question is how to bring this desirable feature of Gauss-Bonnet gravity down to four dimension where Gauss-Bonnet term makes no contribution? 

Another interesting question is of reciprocity between matter and gravity/curvature. As matter produces gravity, gravity should also manifest as matter under some circumstances. The creation of electron-positron pair from radiation was the first demonstration of this principle. The nearest we have come to it in the string theory is the case of AdS/CFT correspondence \cite{mal} where conformal field theory sits on the boundary of the bulk AdS spacetime. In either of these, there is an involvement of a special circumstance, the presence of an external agency like magnetic field for the former and of higher dimensional bulk and lower dimensional boundary for the latter. The interesting question that arises is, could we construct a classical example depicting reciprocity between matter and spacetime curvature. This is precisely what has been accomplished in our recent work \cite{md1,md2,md3}. 

In here we considered an $n(\ge 6)$-dimensional spacetime with Kaluza-Klein split with the topology of ${\ma M}^4 \times {\ma K}^{n-4}$, where ${\ma K}^{n-4}$ is the $(n-4)$-dimensional space of constant curvature with the constant warp factor $r_0$. 
It turns out that the vacuum Einstein-Gauss-Bonnet equation with a cosmological constant $\Lambda$ gets correspondingly split up into two parts \cite{md1}. The four-dimensional part is like the vacuum Einstein equation with a cosmological constant redefined, which has the general solution as Schwarzschild-dS/AdS while the extra dimensional part gives a scalar constraint equation. 
Then the constraint does not permit presence of a mass point and hence the vacuum solution reduces to a dS/AdS spacetime in four dimension. 
However there exists an exceptional case in which four-dimensional equation could be made completely vacuous by prescribing $r_0$ and $\Lambda$ in terms of Gauss-Bonnet parameter $\alpha$ and the curvature of extra dimensional space. 
Then $4$-metric is entirely determined for a spacetime satisfying the null energy condition by the scalar constraint given by the extra dimensional equation and it admits the general solution.

The solution represents a black hole with two horizons and ${\ma M}^4$ asymptotically approximates to Reissner-Nortstr\"om-anti-de~Sitter (RN-AdS) spacetime for positive $\alpha$ in spite of absence of Maxwell field. 
Note that the scalar constraint had prohibited presence of any matter, yet however there had come about a black hole with parameters which asymptotically resemble mass and Maxwell charge. This is what we call the pure gravitational creation, ``matter without matter'', produced by curvature of extra dimensional space through its linkage with $\alpha$ and $\Lambda$. It is purely a classical demonstration of reciprocity 
between matter and gravity/curvature which is the first such example after the classic case of Kaluza-Klein. The solution could also be generalized to include Vaidya null dust \cite{md2}. It could then be envisioned that collapsing null dust in this setting create the black hole from an AdS spacetime. 
   
The paper is organized as follows. In the next section, we discuss classical motivations for higher dimension followed by gravitational action and equation in higher dimension. In section 4, we present the black hole created out of AdS spacetime as a creature of pure curvature establishing the reciprocity between matter and curvature. We end with a discussion.     

\section{Higher dimension}
Higher dimension is natural playground for string theory while we would like to 
argue that even classical considerations also ask for it~\cite{dad}.  It is the universal character of gravity which demanded that it could only be described by curvature of spacetime. Its dynamics is determined by the curvature through the Bianchi identity which on contraction leads to Einstein equation. This equation is valid for dimension $n\ge2$. It turns out that two and three dimensions are not big enough to accommodate free propagation of gravity and so we come to four dimensional spacetime. Thus four dimension is necessary for gravitational dynamics, the question is, is it also sufficient? Since gravitational dynamics now fully resides in curvature, the right question to ask is, does curvature information propagate in higher dimension or not? 

One of the indicators of that would be whether four dimensional curved spacetime is isometrically embeddable in five dimensional flat spacetime which is then completely free of gravity. This will ensure non propagation of gravitational dynamics in higher dimension. It is however well known from differential geometry that an arbitrarily curved $n$ dimensional space requires as many as $n(n-1)/2$ extra dimensions for its flat space embedding. This means four dimensional space would require ten dimensions and in other words gravity can in general penetrate down to ten dimension. In particular, though conformally flat FRW model is embeddable in five dimensional flat while the Schwarzschild requires six dimension. It is interesting to note that the number of extra dimensions required 
is the same as the number of metric degrees of freedom. That is, for each degree of freedom corresponds an extra dimension for embedding. This is perhaps not a coincidence and I believe it is perhaps indicative of some deeper geometric connection between Lorentz group degrees of freedom and isometric embedding. 

Second, gravity is inherently self interactive force. Self interaction can be evaluated only by an iteration process. The first iteration is however already contained in Einstein equation as it includes square of first derivative of the metric. The question is, how do we stop at the first iteration, we should go to next iterations? For the second iteration, we should include square of curvature in the action but which will unfortunately square the second derivative as well and thereby making the equation non quasi-linear. Quasi-linearity is required for the initial value problem prescription and the unique evolution. This requirement singles out Lovelock Lagrangian of which Gauss-Bonnet is the quadratic term. For the second iteration, we should therefore include Gauss-Bonnet term which however makes no contribution in the equation of motion for $n<5$. Hence we have to go to five or higher dimension to physically realize the second iteration of self interaction of gravity. So the self interaction dynamics of gravity naturally demands higher dimension. Next question is, where does this iteration chain finally end? At the second level itself! If the matter is entirely 
confined to $3$-brane, the higher dimensional bulk spacetime is completely free of matter. Then it should be homogeneous and isotropic in space and homogeneous in time and is therefore of constant curvature. It is maximally symmetric (dS/AdS) having zero Weyl curvature which indicates absence of free gravity. Since there is no free gravity to propagate any further, the iteration chain terminates at the second level. What we can therefore say is that as two and three dimensions were not big enough to accommodate free gravity, four dimension is not big enough to fully accommodate the self interaction dynamics. 

Third, we appeal to the principle of overall charge neutrality for a classical field;i.e. total charge being zero. This is respected by electromagnetic field, total electric charge is globally zero. The same must be true for gravity as well. Matter/energy is the charge for gravity which is however unipolar - always positive. Then how to neutralize it? The only way it could be done is that gravitational field must have charge of opposite polarity (negative). That is why gravity has to be attractive. The negative polarity is not localizable like the matter, say a mass point, as it is spread all over the space with the field. If we integrate it all over the space it will perfectly balance positive mass of the particle. However in the local neighborhood around the mass point, there will be over dominance of positive charge, and hence the field should propagate of the $3$-brane in that neighborhood. The off propagation will not however be free as it propagates its past light cone will keep on enclosing more and more of negative charge of the spread out field and thereby diminishing its strength. That is, as field propagates off the brane its strength goes on diminishing. That means it will not be able to penetrate deep enough into the bulk. Since its strength diminishes with propagation, it could be envisioned as having the running coupling for propagation, like the strong force, in extra dimension. Like AdS/CFT correspondence between gravity and strong force, here we have another indication of similar kind of relation between them. This is however an intuitive and qualitative argument which needs to be established quantitatively and rigorously. 

These are the three purely classical motivations for higher dimension \cite{dad,dad1}.

 
\section{Gravity in higher dimension}
We write action for $n$-dimensional spacetime, 
\begin{equation} 
\label{action}
S=\int d^nx\sqrt{-g}\biggl[\frac{1}{2\kappa_n^2}(R-2\Lambda+\alpha{L}_{GB}) \biggr]+S_{\rm matter},
\end{equation}
where $R$ and $\Lambda$ are $n$-dimensional Ricci scalar and the cosmological constant, respectively. 
Further $\kappa_n\equiv\sqrt{8\pi G_n}$, where $G_n$ is $n$-dimensional gravitational constant and $\alpha$ is the Gauss-Bonnet coupling constant.
The Gauss-Bonnet term ${L}_{GB}$ is combination of squares of Ricci scalar, Ricci tensor $R_{\mu\nu}$, and Riemann tensor $R^\mu_{~~\nu\rho\sigma}$ as
\begin{equation}
{L}_{GB} \equiv R^2-4R_{\mu\nu}R^{\mu\nu}+R_{\mu\nu\rho\sigma}R^{\mu\nu\rho\sigma}.
\end{equation}

This type of action is derived in the low-energy limit of heterotic superstring theory~\cite{Gross}.
In that case, $\alpha$ is identified with the inverse string tension and is positive definite. 
We shall therefore take $\alpha \ge 0$.

The gravitational equation following from the action (\ref{action}) is given by 
\begin{equation}
{\ma G}^\mu_{~~\nu} \equiv {G}^\mu_{~~\nu} +\alpha {H}^\mu_{~~\nu} +\Lambda \delta^\mu_{~~\nu}=\kappa_n^2 {T}^\mu_{~~\nu}, \label{beq}
\end{equation}
where 
\begin{eqnarray}
{G}_{\mu\nu}&\equiv&R_{\mu\nu}-{1\over 2}g_{\mu\nu}R,\\
{H}_{\mu\nu}&\equiv&2\Bigl[RR_{\mu\nu}-2R_{\mu\alpha}R^\alpha_{~\nu}-2R^{\alpha\beta}R_{\mu\alpha\nu\beta} \nonumber \\
&&~~~~+R_{\mu}^{~\alpha\beta\gamma}R_{\nu\alpha\beta\gamma}\Bigr]
-{1\over 2}g_{\mu\nu}{L}_{GB}.\label{def-H}
\end{eqnarray}
$T_{\mu\nu}$ is the energy-momentum tensor of the matter field derived from $S_{\rm matter}$ in the action (\ref{action}). 
It is noted that Gauss-Bonnet term makes no contribution in the field equations, i.e. $H_{\mu\nu} \equiv 0$, for $n \le 4$.

We consider the $n$-dimensional spacetime locally homeomorphic to ${\ma M}^{4} \times {\ma K}^{n-4}$ with the metric, $g_{\mu\nu}=\mbox{diag}(g_{AB},r_0^2\gamma_{ab})$, $A,B = 0, \cdots, 3;~a,b = 4, \cdots, n$. Here $g_{AB}$ is an arbitrary Lorentz metric on ${\ma M}^4$, $r_0$ is a constant and $\gamma_{ab}$ is the unit metric on the $(n-4)$-dimensional space of constant curvature ${\ma K}^{n-4}$ with its curvature $ k = \pm 1, 0$. 
Then ${\ma G}^\mu_{~~\nu}$ gets decomposed as follows:
\begin{eqnarray}
{\ma G}^A_{~~B}&=&\biggl[1+\frac{2 k\alpha(n-4)(n-5)}{r_0^2}\biggl]\overset{(4)}{G}{}^A_{~B} \nonumber \\
&&+\biggl[\Lambda-\frac{k(n-4)(n-5)}{2r_0^2} \nonumber \\
&&-\frac{k^2\alpha(n-4)(n-5)(n-6)(n-7)}{2r_0^4}\biggl] \delta^A_{~B},\label{dec1} \\
{\ma G}^a_{~~b}&=&\delta^a_{~~b}\biggl[-\frac12\overset{(4)}{R}+\Lambda-\frac{(n-5)(n-6)k}{2r_0^2} \nonumber \\
&&-\alpha\biggl\{\frac{k(n-5)(n-6)}{r_0^2}\overset{(4)}{R}+\frac12 \overset{(4)}{L}_{GB} \nonumber \\
&&+\frac{(n-5)(n-6)(n-7)(n-8)k^2}{2r_0^4}\biggl\}\biggl],\label{dec2}
\end{eqnarray}
where the superscript $(4)$ means the geometrical quantity on ${\ma M}^4$.  

The decomposition immediately leads to a general result in terms of the following no-go theorem on ${\ma M}^4$: 
\begin{The} 
\label{the:1}
If (i) $r_0^2=-2k\alpha(n-4)(n-5)$ and (ii) $\alpha\Lambda = -(n^2-5n-2)/[8(n-4)(n-5)]$, then  ${\ma G}^A_{~~B} = 0$ for $n \ge 6$ and $k$ and $\Lambda$ being non-zero.
\end{The}
The proof simply follows from substitution of the conditions (i) and (ii) in  Eq.~(\ref{dec1}). 
As a corollary, it states that ${\ma M}^4$ cannot harbor any matter/energy distribution unless at least one of the conditions (i) and (ii) is violated. 

These conditions also imply for $\alpha > 0$, $k = -1$ and $\Lambda < 0$. 
Hereafter we set $k = -1$ and obtain the vacuum solution with $T_{\mu\nu} = 0$ satisfying the conditions (i) and (ii).
The governing equation is then a single scalar equation on ${\ma M}^4$, $\ma G^{a}_{~~b} = 0 $.

\section{Black hole from AdS}
In here with the parameters given in Theorem 1, we shall solve the constraint equation which will read as follows:
\begin{equation}
-\frac{n-2}{2}R-\frac{(n-4)\alpha}{2}L_{GB}+n\Lambda=\kappa_n^2 T = 0 \label{beq-t}
\end{equation}
 
Note that this is a trace zero equation in four dimension and hence it will with the null energy condition ($T_{AB}k^Ak^B = 0, k^Ak_A = 0$) solve to give a Maxwell-like distribution. This is because it is precisely the characterization of Maxwell field in four dimension. The occurrence of Maxwell-like charge in solutions of this equation should not therefore be surprising. We shall now obtain the analogues of Schwarzschild and Vaidya solutions.

\subsection{Schwarzschild like solution}
We seek a static solution for a point mass with the metric on ${\ma M}^4$ reading as:
\begin{equation}
g_{AB}dx^Adx^B=-f(r)dt^2+\frac{1}{f(r)}dr^2+r^2d\Sigma^2, \label{NdS2} 
\end{equation}
where $d\Sigma^2$ is the unit metric on ${\ma K}^2$.

It is clear that Eq.~(\ref{dec1}) is equivalent to Einstein equation with redefined $\Lambda$ and it will admit Schwarzschid-dS/AdS as the general solution. When it is put into the constraint, Eq.~(\ref{dec2}), which would demand $M=0$ reducing the solution to dS/AdS. However there exists a special case indicated by the Theorem 1, which makes the four dimensional equation completely vacuous leaving the metric entirely free and arbitrary.  
Then the constraint equation, $\ma G^{a}_{~~b} = 0$ given by Eq. (\ref{beq-t}), yields the general solution for the function $f(r)$: 
\begin{eqnarray}f(r)&=&1+\frac{r^2}{2(n-4)\alpha}\biggl[1\mp\biggl\{1-\frac{2n-11}{3(n-5)} \nonumber \\
&&+\frac{4(n-4)^2\alpha^{3/2}M}{r^3}-\frac{4(n-4)^2\alpha^2q}{r^4}\biggl\}^{1/2}\biggl], \label{special} 
\end{eqnarray}
where $M$ and $q$ are arbitrary dimensionless constants. 
The solution does not have the general relativistic limit $\alpha \to 0$.
There are two branches of the solution indicated by sign in front of the square root in Eq.~(\ref{special}), and it is the minus branch which gives the black hole.

The function $f(r)$ is expanded for $r \to \infty$ as
\begin{eqnarray}
f(r)&\approx& 1\mp \frac{\alpha^{1/2} M\sqrt{3(n-4)(n-5)}}{r} \nonumber \\
&&\pm \frac{\alpha q\sqrt{3(n-4)(n-5)}}{r^2} \nonumber \\
&&+\frac{r^2}{2(n-4)\alpha}\left(1\mp\sqrt{\frac{n-4}{3(n-5)}}\right). \label{infty}
\end{eqnarray}
This is the same as the Reissner-Nordstr\"om-anti-de~Sitter (AdS) spacetime in spite of the absence of the Maxwell field. This suggests that $M$ is 
the mass of the central object and $q$ is the charge-like new parameter. On the other hand, $f(r=0) = 1\mp\sqrt{-q}$ implying that the metric remains finite and regular everywhere and so does its derivative $f'$. It is clear that $q$ must be negative which will also ward off any branch singularity by not letting the term under the square-root become negative. It is a new Kaluza-Klein gravitational charge like the Weyl charge of the black hole on the brane \cite{dmpr}. As for the Gauss-Bonnet black hole \cite{boul,dad} in five dimension, singularity is also weakened as Riemann curvature diverges as $1/r^2$ instead of $1/r^3$ for Schwarzschild. As envisaged in the Introduction, we have thus brought down the desirable features of Gauss-Bonnet gravity down to four dimension by this novel construction. 

More importantly, this black hole is a pure curvature creation as the equation completely prohibits presence of any matter distribution including a mass point delta distribution and the spacetime emerges entirely as the solution of the extra dimensional scalar equation. It is the prescription of the parameters given in Theorem 1 which is responsible for the absence of matter support ${\ma M}^4$.  

\subsection{Vaidya-like solution}
It is well known that Schwarzschild spacetime could be made to radiate null (Vaidya) radiation by transforming the metric into retarded/advanced time coordinate and then making mass parameter function of the time coordinate. 
It is interesting to note that the same procedure also works here. 
This solution~(\ref{special}) can thus be generalized to include Vaidya radiation and it would be given by
\begin{widetext}
\begin{eqnarray}
g_{AB}dx^Adx^B&=&-{\tilde f}(v,r)dv^2+2dvdr+r^2d\Sigma^2, \label{NdS-d} \\
{\tilde f}(v,r)&\equiv&1+\frac{r^2}{2(n-4)\alpha}\biggl[1\mp\biggl\{1-\frac{2(n-5)(2n-11)}{6(n-5)^2} \nonumber \\
&&~~~~~~~~~+\frac{4(n-4)^2\alpha^{3/2}{\tilde M}(v)}{r^3}-\frac{4(n-4)^2\alpha^2{\tilde q}(v)}{r^4}\biggl\}^{1/2}\biggl], \label{special-d}
\end{eqnarray}
\end{widetext}
where ${\tilde M}(v)$ and ${\tilde q}(v)$ are arbitrary functions. This is the solution of Eq. (\ref{beq-t}) which shows that the usual procedure of making Schwarzschild particle radiate null dust works in this setting as well.
All this works because trace of null dust stress tensor vanishes and these solutions follow from solving the vanishing trace equation (\ref{beq-t}).

\begin{figure}[tbp]
\includegraphics[width=.70\linewidth]{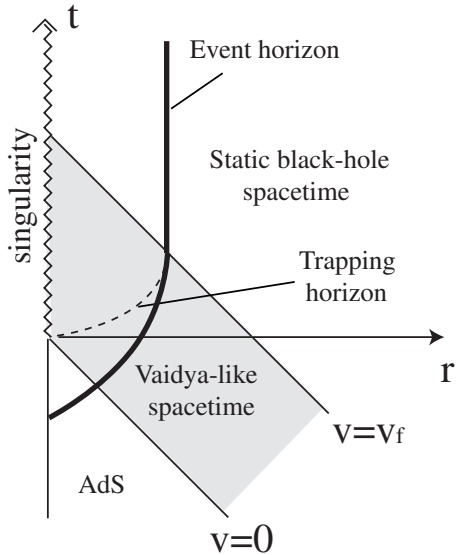}
\caption{
A schematic diagram of the black-hole formation from the AdS spacetime without any matter field.
Here the AdS spacetime for $v<0$ is joined to the static black-hole spacetime (\ref{special}) for $v>v_{\rm f}$ by way of the Vaidya-like spacetime (\ref{NdS-d}).
We set $\Theta=0$, $M(v)=m_0v$ and $q(v)=-1$, where $m_0$ is a positive constant.
A singularity is formed at $v=r=0$ and develops.
}
\label{Fig1}
\end{figure}

This solution manifests gravitational creation of an ingoing charged null dust as another complete example of ``matter without matter''.
Using this solution and the solution~(\ref{special}), we can construct completely vacuum spacetime representing the formation of a black hole from an AdS spacetime by gravitational collapse of a gravitationally created charged null dust (See Fig.~\ref{Fig1}).

\section{Discussion}
In this paper, we had set out to do two things: (i) to bring desirable features of Gauss-Bonnet gravity down to four dimension and (ii) to establish reciprocity between matter and curvature;i.e. to create a black hole out of a pure curvature. Both these aims have been accomplished. 

New black hole formed out of coupling of extra dimensional curvature with $\alpha$ and $\Lambda$ inherits the Gauss-Bonnet features of regularity of metric and its first derivative everywhere and weakening of the central singularity as curvatures diverge only as $1/r^2$. Kaluza-Klein splitting of spacetime gives rise to additional gravitational charge that resembles Maxwell charge as the equation determining the spacetime is of the vanishing trace which is the characteristic of Maxwell field with the null energy condition in four dimension. Since the solution could be generalized in the usual manner to Vaidya analogue, it could be envisioned that the black hole could be formed out of an AdS spacetime by collapse of radially infalling null dust. This is perhaps the only way such a black hole could be formed as it cannot have a static interior because four dimensional spacetime cannot harbor any matter distribution because of the specific prescription of the parameters. 

Here we have demonstrated the creation of trace free matter in a very novel 
way, however the question of formation of matter with non zero trace remains open. That is what we would like to pursue in future. The further details could be had from Refs \cite{md1,md2,md3}.

\acknowledgments
I wish to thank Hideki Maeda for the fruitful collaboration resulting in 
Refs.~\cite{md1,md2} on which this talk is based except for most of the discussion in Sec. 2.   


\end{document}